\documentclass[prl,twocolumn,showpacs]{revtex4}
\usepackage{amsmath,amssymb,xspace,graphicx}

\newcommand{\rmc}{{\mathrm c}}
\newcommand{\rmd}{{\mathrm d}}
\newcommand{\rmi}{{\mathrm i}}
\newcommand{\rmb}{{\mathrm B}}
\newcommand{\rmf}{{\mathrm F}}
\newcommand{\rml}{{\mathrm L}}
\newcommand{\rmr}{{\mathrm R}}

\newcommand{\bk}{\boldsymbol{k}}

\newcommand{\pal}{\partial_l}

\newcommand{\ie}{i.e.\@\xspace}

\begin{document}
\title{Effects of non-magnetic impurities on spin-fluctuations induced
superconductivity}

\author{S. Dusuel}
\address{Institut f\"ur Theoretische Physik, Universit\"at zu K\"oln,
Z\"ulpicher Str. 77, D-50937 K\"oln, Germany}

\author{D. Zanchi}
\address{Laboratoire de Physique Th\'eorique et Hautes \'Energies, 2 Place Jussieu,\\ 75252 Paris C\'edex 05, France.}

\date{\today}


\begin{abstract}
  
  We study the effects of non-magnetic impurities on the phase diagram of 
  a system of interacting electrons with a flat Fermi surface.
  The one-loop Wilsonian renormalization group flow of the angle dependent
  diffusion function $D(\theta_1,\theta_2,\theta_3)$
  and interaction $U(\theta_1,\theta_2,\theta_3)$
  determines the critical temperature  and the nature of the 
  low temperature state.
  As  the imperfect nesting increases the critical temperature decreases 
  and the low temperature phase changes from the spin-density wave (SDW) 
  to the d-wave superconductivity (dSC) and finally,
  for bad nesting, to the random antiferromagnetic state (RAF).
  Both SDW and dSC phases are affected by disorder.
  The pair breaking depends on the imperfect nesting and is the most efficient
  when the critical temperature for superconductivity is maximal. 
  
\end{abstract}


\maketitle


Interesting and new physics appears when density wave (DW) and 
superconducting (SC)
correlations interfere with each other and with the disorder.
One finds realizations of such interferences in correlated 
electronic systems whose Fermi surface (FS) has at least 
imperfect nesting properties.
The Bechgaard salts (TMTSF)$_2$X and the Fabre  salts (TMTTF)$_2$X 
have two open quasi-one-dimensional Fermi sheets
responsible for DW correlations and, very probably, for 
superconductivity.\cite{Ishiguro} 
In these compounds the disorder strength can  be tuned
by the cooling speed or by x-ray irradiation,
while the imperfect nesting is controlled either by
chemical structure or by pressure.
Another example
of correlated system with important  disorder effects 
are the two-dimensional (2D) organic compounds from the BEDT-TTF (ET) 
family.\cite{Lang03,Ishiguro}
All ETs 
have a metal-insulator and/or a DW transition 
in the phase diagram. They have two FSs with several more or less 
perfect inter- and intra-band nesting wave vectors. Their superconductivity 
is induced by the DW fluctuations. 
In the high-$T_\rmc$ superconductors
the pseudogap and the 
superconducting states are both related to the antiferromagnetic (commensurate 
SDW) insulating phase.
The effects of disorder on the high-$T_c$ 
superconductivity has been widely studied experimentally and 
theoretically.\cite{Maki98,Hussey02} 
It has been shown by a number of methods, and in 
particular by the angle resolved renormalization group (RG)
that this interplay between the DW, the pseudogap and the SC exists in 
models with at least
flat segments of the FS or with nested van 
Hove singularities.\cite{Zheleznyak97,Zanchi00,Furukawa98} 
However, no attempt has been made to implement both disorder and correlations
in the N-patch RG, which would provide a perturbative, but non-biased and 
controlled way for constructing the phase diagram.

In this letter we report a perturbative angle-resolved 
(N-patch) RG theory for the model consisting of two flat FS segments
 in regard. 
The imperfect nesting parameter is treated approximately, by introducing
a cut-off $\epsilon$ in the Peierls channel.
For the pure case this problem was solved in the 
Parquet formulation by Zheleznyak et al \cite{Zheleznyak97} and later closely 
reexamined in the field-theoretical 
RG language.\cite{Dusuel02}
While previous theories in more than one dimension (1D) were based on the
assumption of one
dominant order parameter,\cite{Abrikosov,Lee85,Belitz94} we take into account
in a non-prejudiced way both Cooper and Peierls correlations together with the
diffusion processes, generalizing to 2D the 1D theory of
Giamarchi and Schulz.\cite{Giamarchi88}

\begin{figure}[t]
  \centering
  (a)\includegraphics[width=2.cm]{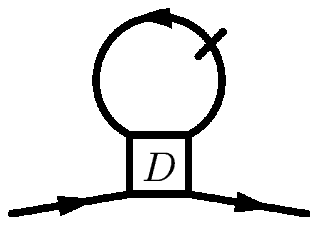}\hspace{0.2cm}(b)\includegraphics[width=2.3cm]{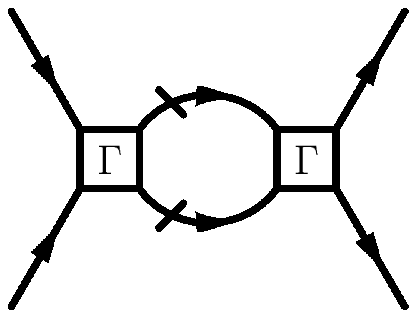}\hspace{0.2cm}(c)\includegraphics[width=2.3cm]{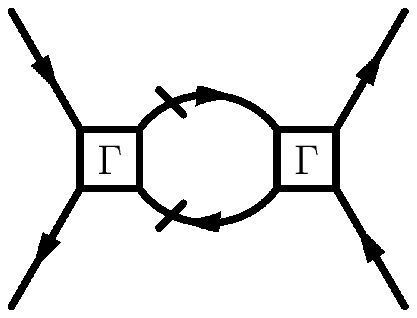}
  \caption{Self-energy (a), particle-hole (b) and particle-particle (c)
    graphs taken into account. $\Gamma$ symbolises both interactions $U$
    and disorder $D$. Barred propagators are on-shell at $\Lambda$.}
  \label{fig:feynman_graphs}
\end{figure}
\begin{figure}[t]
  \centering
  \includegraphics[height=4.5cm]{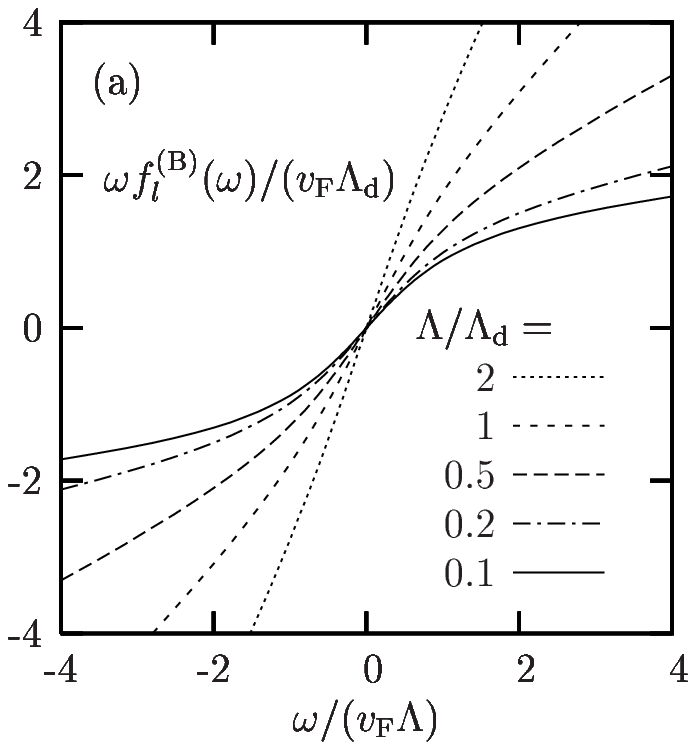}\hspace{0.3cm}\includegraphics[height=4.5cm]{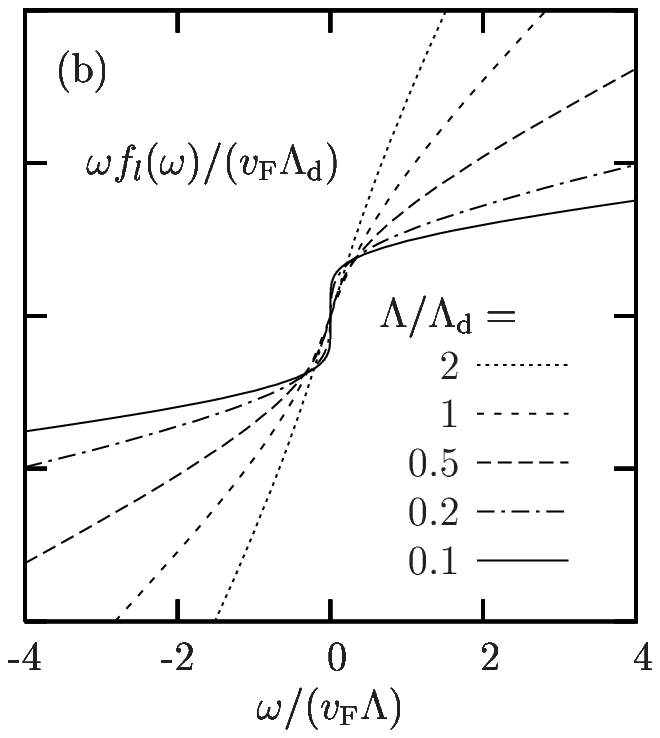}
  \caption{Progressive building of the $\omega$-dependence of $f_l(\omega)$ as
    modes are eliminated. (a) is the Born approximation
    $f^{(\mathrm{B})}_l(\omega)$ and (b) is the exact solution of
    Eq.~(\ref{eq:RG_eq_f})
  }
  \label{fig:f_w}
\end{figure}

We first consider a non-interacting isotropic disordered electronic system
with bare dispersion $\xi(\bk)$.
We constrain the momenta to lie within the shell of $\pm \Lambda$ 
around the Fermi surface and follow the Wilsonian one-loop flow starting 
from $\Lambda_0$.
At the scale $l=\ln (\Lambda_0/\Lambda)$ the propagator
has the form 
\begin{equation}
  \label{eq:prop_ren}
  \mathcal{G}_l^{-1}(\bk,\omega)=\rmi\omega f_l(\omega) - \xi(\bk),
\end{equation}
where $f_l$ is a flowing {\em function} of $\omega$.
Considering only the self-energy correction due to the disorder, see
Fig.~\ref{fig:feynman_graphs}(a) 
and discarding for simplicity one-loop graphs renormalizing the 
diffusion $D$
(Figs.~\ref{fig:feynman_graphs}(b) and (c)), we find
\begin{equation}
  \label{eq:RG_eq_f}
  \frac{\rmd f_l(\omega)}{\rmd \Lambda}=-\frac{v_\rmf}{\pi\tau }  \, 
\frac{f_l(\omega)}{(v_\rmf\Lambda)^2+[\omega f_l(\omega)]^2}, 
\mbox{ with } f_{l=0}(\omega)=1\; ,
\end{equation}
where $\tau=v_\rmf/2D$ is the elastic scattering mean free time. 
The Born approximation 
$f^{(\rmb)}=1+1/(2|\omega|\tau)$
is recovered in the limit
$\Lambda\to 0$, when setting $f_l(\omega)$ to its bare value $f_l=1$, in the RHS of
Eq.~(\ref{eq:RG_eq_f}).
In Fig.~\ref{fig:f_w} we see how the $\omega$-dependence of $f$ is
constructed successively by mode elimination.
For any finite $\Lambda$, the one-particle propagator is a 
{\em regular} function of 
frequency. 
The characteristic momentum scale for disorder is 
$\Lambda_\rmd=2D/(\pi v_\rmf^2)$.
In the Born approximation, and for 
$\Lambda/\Lambda_\rmd\ll 1$, the slope of
$\omega f_l^{(\rmb)}(\omega)$ at the origin is
$f_l^{(\rmb)}(0)\simeq {\Lambda}_\rmd/{\Lambda}$. 
The exact solution of
Eq.~(\ref{eq:RG_eq_f}) is much steeper since its slope is
$f_l(0)\simeq\exp(\Lambda_\rmd/\Lambda)$.


We now consider the disordered and interacting system whose FS consists
of two flat segments in regard.
We introduce left (L) and right (R) angle-resolved movers, and
decompose the momentum $\bk$ into one component $k$ orthogonal to the FS and
the other $k_\|$ parallel to the FS. The angular position on the FS is denoted
by $\theta=(\rmr/\rml,k_\|)$.
We will assume the Fermi velocities do not depend on $\theta$.
In the one-loop approximation, the one-particle propagator at scale $l$ is
$\mathcal{G}_l^{-1}(k,\omega,\theta)=\rmi\omega f_l(\omega,\theta) - 
v_\rmf k$.
A complete description of the disordered phases would require a fully
functional RG, taking the whole frequency dependence of the propagator into
account.
Here we consider the problem from the weakly coupled Fermi liquid side. We
want to study how the model is driven away from its Fermi liquid fixed point
by the combined effects of weak interactions and disorder.
For this purpose the zero-order scaling arguments \cite{Shankar94} allow us
to keep only terms linear in energy in the non-interacting part of
the replicated (see e. g. \cite{Belitz94}) effective action, which then reads
\begin{equation}
  \label{eq:S0}
  S_0=-\int^{\Lambda}\frac{\rmd X}{2\pi}
  \Big[\rmi\omega Z_l^{-1}(\theta) - v_\rmf k\Big]
  \bar\psi(X)\psi(X).
\end{equation}
$X=(k,\omega,\theta,\sigma,\alpha)$ is a multiple index.
The integral over $X$ is a shorthand notation for
integrals over $k$, $\omega$ and $k_\|$ and discrete summations over
the replica index $\alpha$, spin index $\sigma$ and the left-right index.
In Eq.~(\ref{eq:S0}) the quantity $Z_l^{-1}(\theta)
\equiv f_l(\omega=0,\theta)$. It
should not be confused 
with the one-particle weight, discussed
in \cite{Zanchi01}. In fact, the effective action (\ref{eq:S0})
can be interpreted as the marginal part of the diffusion pole as ``seen'' by 
the electrons at the RG scale $l$.
%
%
%
%
The interaction and the disorder parts of the replicated effective action read
\begin{widetext}
  \begin{eqnarray}
    \label{eq:Sint}
    &&S_\mathrm{int}=2\pi v_\rmf
    \int^{\Lambda}\frac{\rmd X_1\ldots\rmd X_4}{(2\pi)^4}
    [2\pi\delta(4+3-2-1)]\delta_{\alpha_1,\alpha_4}\delta_{\alpha_2,\alpha_3}
    \delta_{\alpha_1,\alpha_3}U_l(\theta_1,\theta_2,\theta_3)
    \bar\psi_4\bar\psi_3\psi_2\psi_1,\\
    \label{eq:Sdis}
    &&S_\mathrm{dis}=2\pi v_\rmf\frac{v_\rmf\Lambda_0}{2}
    \int^{\Lambda}\frac{\rmd X_1\ldots\rmd X_4}{(2\pi)^4}
    [2\pi\delta(4+3-2-1)]\delta_{\alpha_1,\alpha_4}\delta_{\alpha_2,\alpha_3}[2\pi\delta(\omega_1-\omega_4)]D_l(\theta_1,\theta_2,\theta_3)
    \bar\psi_4\bar\psi_3\psi_2\psi_1.
  \end{eqnarray}
\end{widetext}
We have denoted $\psi_i=\psi(X_i)$.
The $\delta(4+3-2-1)$ function ensures momentum and frequency conservation.
$U$ couples electrons in the same replica only.
We keep only the angle dependence of $U$, and drop momentum and
frequency dependence since we restrict ourselves to the marginal part
of the interaction. We do not take Umklapp processes into account 
because we suppose that the system is away from half-filling.
The disorder part of the effective action is the generalisation of
the one used by Giamarchi and Schulz.\cite{Giamarchi88}
Two angle dependent diffusion functions $D_{\eta}$ (forward)
and $D_{\xi}$ (backward) are gathered in a single diffusion function
$D(\theta_1,\theta_2,\theta_3)$.
The disorder couples electrons from two different replicas and in the same
replica. 
The $\delta(\omega_1-\omega_4)$ function ensures the elastic character of the
diffusion function.
We keep only the relevant part of the diffusion function and neglect all
marginal contributions, which are linear in  momenta perpendicular to the FS
and in frequencies.


We derive one-loop RG flow
equations by successive integration of fast modes.
The Feynman graphs generated by infinitesimal mode elimination
are shown on Fig.~\ref{fig:feynman_graphs}.
Schematically, the angle resolved  flow equations  read
\begin{eqnarray}
  \label{eq:RG_flow_eq}
  &&\pal \widetilde{U}_l(\theta_1,\theta_2,\theta_3)=
  \beta_U\big[\widetilde{U}_l,\widetilde{D}_l\big]
  (\theta_1,\theta_2,\theta_3),\\
  &&\pal \widetilde{D}_l(\theta_1,\theta_2,\theta_3)=
  \Big(\widetilde{D}_l+\beta_D\big[\widetilde{U}_l,\widetilde{D}_l\big]\Big)
  (\theta_1,\theta_2,\theta_3).\nonumber\quad
\end{eqnarray}
with $\widetilde{U}_l(\theta_1,\theta_2,\theta_3)=\sqrt{Z_1Z_2Z_3Z_4}\; 
U_l(\theta_1,\theta_2,\theta_3)$ and 
$\widetilde{D}_l(\theta_1,\theta_2,\theta_3)=\exp(l)\sqrt{Z_1Z_2Z_3Z_4}\; 
D_l(\theta_1,\theta_2,\theta_3)$, where $Z_i=Z_l(\theta_i)$.
Note that the equation for the disorder contains a linear term in $D$,
explicitly showing the relevance of the disorder.
The  $\beta$-functions are homogeneous quadratic functions in $\tilde{U}$ and 
$\tilde{D}$.
The disorder is not renormalized by a term quadratic in $\tilde{U}$
and the interactions are not renormalized by a term quadratic in $\tilde{D}$.
Indeed, starting from a pure situation, interactions alone cannot induce
disorder and
starting from a non-interacting disordered situation,
elastic couplings (the disorder) cannot induce inelastic couplings
(the interactions).
We checked our RG equation in two limits.
In the 1D limit, neglecting the contributions quadratic in
the disorder, we recover the (weak coupling) equations
of Giamarchi and Schulz.\cite{Giamarchi88}
The effect of disorder on a conventional $s$-wave superconductor is 
obtained from our equations when dropping the particle-hole contributions and
using initial conditions of the attractive Hubbard type.
In this case we recover Anderson's theorem.\cite{Anderson59}

\begin{figure}[t]
  \centering
  \includegraphics[width=8cm]{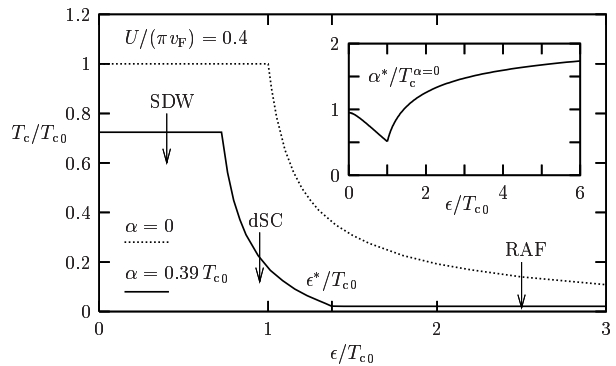}
  \caption{Effect of disorder on the phase diagram.
    Inset shows the critical pair breaking parameter.
    The dip at $\epsilon=T_{c0}$ corresponds to the 
    maximal SC critical temperature for the pure case.
  }
  \label{fig:ph_diag}
\end{figure}
\begin{figure*}[t]
  \centering
  \includegraphics[height=5cm]{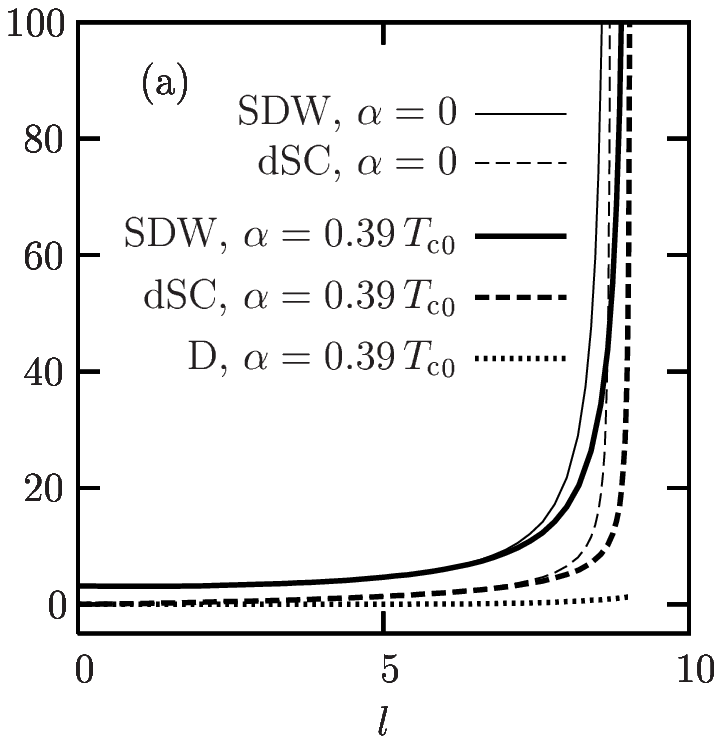}
  \includegraphics[height=5cm]{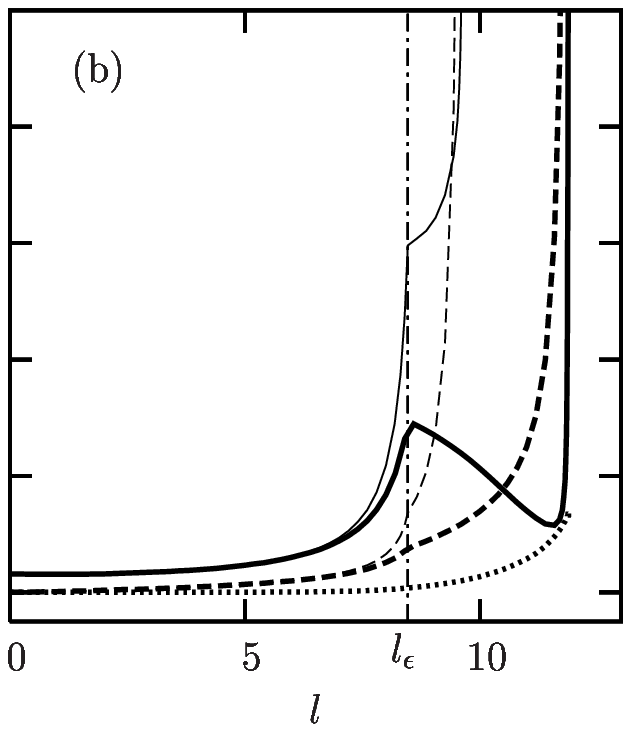}
  \includegraphics[height=5cm]{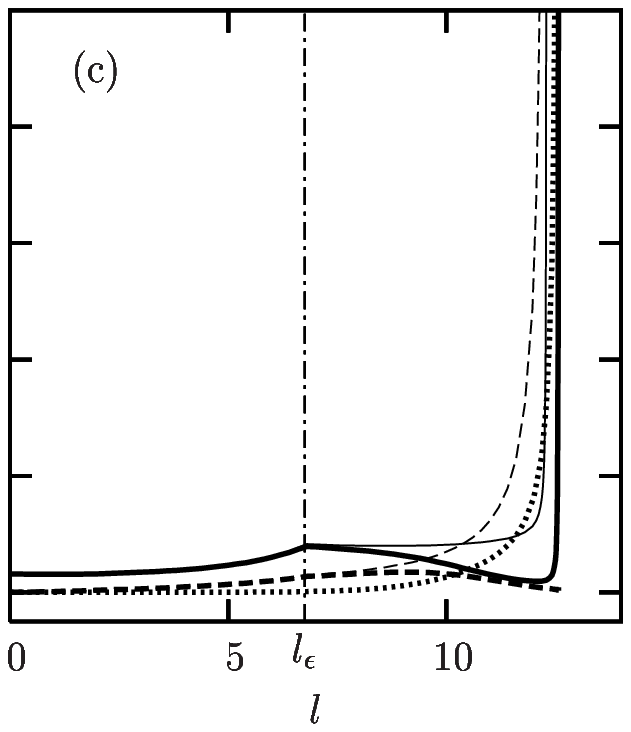}
  \caption{Flow of dominant components of coupling U and diffusion D, for 
    (a) perfect nesting,
    (b) imperfect nesting ($l_\epsilon=\ln(\Lambda_0/\epsilon)=\simeq 8.5$) and (c) bad nesting ($l_\epsilon\simeq 6.8$).}
  \label{fig:flow_coup_total}
\end{figure*}

After discretization of angular variables, we have integrated 
numerically the equations (\ref{eq:RG_flow_eq}). The robustness of our results 
upon increasing number of patches is satisfactory above $N=16$ patches
per side of the FS.
The dependence of the critical scale on the imperfect nesting gives  
the phase diagram.  
For the pure case, \ie the pair-breaking parameter $\alpha=1/\tau=0$,
we reproduce the result of Zheleznyak et al, 
shown in dotted line on Fig.~\ref{fig:ph_diag}. 
As one can conclude from the flow of dominant eigenvalues of
the coupling function $U$, shown in thin lines on
Fig.~\ref{fig:flow_coup_total},
the  part of the phase diagram with $T_\rmc>\epsilon$ shows a transition to
the SDW state with enhanced dSC fluctuations, 
while the $T_\rmc<\epsilon$ part corresponds to the onset of $d$-wave
superconductivity. 
We now introduce a finite disorder via the 
pair breaking parameter $\alpha=2D_0/v_\rmf=0.39T_{c0}$, 
$D_0$ being the bare diffusion
constant. The SDW and dSC phases are both affected by disorder.
For $\epsilon$ grater than a disorder dependent critical imperfect nesting 
$\epsilon ^*$
the dSC phase
is completely suppressed and a new low temperature state appears.
The signatures of all three phases are visible on Fig.~\ref{fig:flow_coup_total}. 
At perfect and good nesting, \ie as long as $\epsilon < T_\rmc$, the flow 
near the divergence is almost the same as for the pure case: the SDW 
correlations are dominant while the dSC ones are also divergent but weaker.
The disorder remains finite at the critical scale.
For moderate imperfect nesting, $T_\rmc<\epsilon<\epsilon^*$, the dSC becomes
the dominant instability because the particle-hole logarithms are cut.
The disorder remains finite, but is stronger than in the perfect nesting case.
Finally, at very bad nesting, $\epsilon>\epsilon^*$ the most diverging
``coupling'' is the disorder. The dSC is irrelevant, while the SDW amplitude 
shows the intriguing re-appearance at the very last moment before divergence.
This divergence is induced by the particle-particle $U$-$D$ terms in the
renormalization of the interaction $U$. Its very small residue reflects the 
fact that the SDW interactions which are
 driven by disorder are only these with 
total momentum equal zero. Consequently  only a small subset of all 
SDW amplitudes diverges. Altogether the flow 
in Fig.~\ref{fig:flow_coup_total}(c)
allows us to characterise rather precisely the corresponding low 
temperature state~: it is a charge-localized state with weakly enhanced SDW 
correlations. We expect that the properties of this phase are similar to the
1D random antiferromagnetic (RAF) phase, 
discussed in \cite{Giamarchi88}.

We have calculated the critical pair breaking $\alpha^*$
as a function of the imperfect nesting parameter $\epsilon$.
The result is shown in the inset of Fig.~\ref{fig:ph_diag}. 
As long as the flow is entirely in the parquet regime the pair breaking
efficiency increases with imperfect nesting, i.e.
$\alpha ^*/T_\rmc^{\alpha=0}$ decreases.
At the point $\epsilon=T_\rmc$, where the flow starts to be  
affected by the imperfect nesting, $\alpha^*/T_\rmc^{\alpha=0}$
reaches its minimum.
This is the point of the most efficient pair breaking. It corresponds to
the maximal value of the dSC critical temperature.
As $\epsilon$ increases further the pair breaking gets progressively less 
efficient. The dependence of the pair breaking on the imperfect nesting is 
a very consequence of the interplay between Peierls, Cooper and localization
tendencies through the angular dependence of corresponding amplitudes.
In the Abrikosov-Gor'kov theory,\cite{Abrikosov61,Sun95}
where particle-hole logarithms are neglected, and a dSC order parameter is
phenomenologically assumed, the ratio
$\alpha ^*/T_\rmc^{\alpha=0}\simeq 0.88$ is
independent of nesting. 

In conclusion, we have used the one-loop N-patch renormalization group to 
compute the non-magnetic disorder effects on a system with spin-density
wave fluctuations induced d-wave superconductivity. 
The pair breaking is the most efficient when the $T_\rmc$ for superconductivity 
is maximal.

\begin{acknowledgments}
We are grateful to
C. Bourbonnais, D. Carpentier, B. Dou\c{c}ot, N. Dupuis, T. Giamarchi, K. Maki
and E. Orignac for discussions and comments. S.~D. also wishes to thank
Pr. E. M\"uller-Hartmann's group for hospitality and discussions.
\end{acknowledgments}

\end{document}